\documentclass[12pt]{article}
\usepackage{a4wide}

\usepackage[dvipdfmx,hiresbb]{graphicx} 
\usepackage{mediabb}                    

\usepackage{amsmath,amssymb}
\usepackage{color}
\usepackage[bookmarks,bookmarksnumbered]{hyperref}
\usepackage{cellspace}
\usepackage{tikz}
\usepackage{mathtools}
\usepackage{cite}
\usetikzlibrary{fadings}
\usetikzlibrary{patterns}
\usetikzlibrary{shadows.blur}
\usetikzlibrary{shapes}
\newcommand{\aln}[1]{\begin{align}#1\end{align}}
\newcommand{\nn}{\nonumber\\}

\begin{document}
\title{\vspace{-3cm}
\vbox{
\baselineskip 14pt
\hfill \hbox{\normalsize 
}}  \vskip 1cm
\bf \Large Field Theory for Superconducting Branes and \\
Generalized Particle-Vortex Duality 
\vskip 0.5cm
}
\author{
Kiyoharu Kawana\thanks{E-mail: \tt kkiyoharu@kias.re.kr}
\bigskip\\
\normalsize
\it 
School of Physics, Korean Institute for Advanced Study, Seoul 02455, Korea
\smallskip
}
\date{\today}

\maketitle   
\begin{abstract} 

We propose a field theory of closed $p$-brane $C_p^{}$ interacting with  a $(p+1)$-form gauge field $A_{p+1}^{}$. 
This is a generalization of the Ginzburg-Landau theory (Abelian-Higgs model) for superconducting particles to higher-dimensional superconducting  branes. 
A higher-form gauge invariant action is constructed by utilizing the Area derivative, which is a higher-dimensional generalization of the ordinary derivative.  
We find that the fundamental phenomena of superconductivity, such as the Meisser effect, topological defects, topological order, are naturally extended in the brane-field theory.  
We explicitly construct a topologically non-trivial static configuration that is characterized by the first homotopy group. 
Then, we calculate the low-energy effective theory in the presence of the topological defect and find that it is described by a BF-type topological field theory coupled with the world-volume of the topological defect. 
We also conjecture an infra-red duality between the superconducting brane-field model and dual brane-field model with a global $\mathrm{U}(1)$ higher-form symmetry as a generalization of the Particle-Vortex duality.

\end{abstract} 

\setcounter{page}{1} 

\newpage  

\tableofcontents   

\newpage  

\section{Introduction}\label{Sec:intro}

Ginzburg-Landau (GL) theory provides  a comprehensive and effective framework to describe 
superconducting phases of various matters~\cite{Ginzburg:1950sr,Landau:1937obd,landau2013statistical}.  
For a given gauge group $G$, the condensate of Cooper pair of fermions is represented by a complex scalar field $\phi(X)$, 
which couples to 
%
the gauge field $A_1^{}=A_\mu^{}(X)dX^\mu$ with a non-trivial representation of $G$. 
The phase transition between Coulomb and superconducting phases is described by the spontaneous breaking of the gauge symmetry, where order-parameter is given by the vacuum expectation value (VEV) of $\phi(X)$. 
In the superconducting phase, the gauge bosons become massive due to the Higgs mechanism, which results in the expulsion of magnetic field, i.e. the Meissner effect. 

In modern physics, however, it is more appropriate to understand superconducting transitions from the point of view of spontaneous breaking of higher-form symmetry~\cite{Gaiotto:2014kfa,Kapustin:2005py,Pantev:2005zs,Nussinov:2009zz,Banks:2010zn,Kapustin:2013uxa,Aharony:2013hda,Kapustin:2014gua,Gaiotto:2017yup,Hirono:2018fjr,Hidaka:2019jtv,Hidaka:2022blq}. 
Apparently, the charged field $\phi(X)$ is not a gauge invariant quantity, which motivates us to develop a gauge invariant order-parameter. 
Here, higher-form symmetry comes into play. 
The ordinary GL theory has a magnetic $(d-3)$-form symmetry, whose charged object is the $(d-3)$-dimensional 't Hooft operator in a $d$-dimensional spacetime. 
In the Coulomb phase, the magnetic $(d-3)$-form symmetry is spontaneously broken, indicated by the perimeter law $\sim e^{-L^{d-3}}$ of the 't Hooft operator. 
(Here, $L$ is a typical size of the operator.)
In particular, the massless photon in the Coulomb phase is interpreted as  Nambu-Goldstone mode for the broken magnetic symmetry. 
On the other hand, the magnetic $(d-3)$-form symmetry is unbroken in the superconducting phase, which implies that the 't Hooft operator indicates  the area law $\sim e^{-L^{d-2}}$ due to the confinement of the magnetic flux.  
%
%
%

Despite the great successes of the GL paradigm, it should be noted that it only describes phase transitions of interacting particles, i.e. $0$-dimensional objects in space.    
%
%
In modern physics, however, it is well recognized that the notion of particle is no longer the most fundamental, but it is one of extended objects in spacetime, such as 
%
%
strings (vortices), Wilson ('t Hooft) loops, domain walls, $p$-branes, and more.  
%
%
Correspondingly, the concept of symmetry has also been generalized, and is known as generalized symmetry~\cite{Gaiotto:2014kfa,McGreevy:2022oyu,Gomes:2023ahz,Schafer-Nameki:2023jdn,Brennan:2023mmt,Bhardwaj:2023kri,Luo:2023ive,Shao:2023gho} in recent decades, which includes higher-group~\cite{Sharpe:2015mja,Tachikawa:2017gyf,Cordova:2018cvg,Benini:2018reh,Tanizaki:2019rbk,DelZotto:2020sop,Hidaka:2020iaz,Hidaka:2020izy,Brennan:2020ehu,Hidaka:2021mml,Hidaka:2021kkf,Apruzzi:2021mlh,Barkeshli:2022edm,Nakajima:2022feg,Radenkovic:2022qkd,Bhardwaj:2022scy,Kan:2023yhz} and non-invertible symmetry~\cite{Bhardwaj:2017xup,Chang:2018iay,Ji:2019jhk,Komargodski:2020mxz,Nguyen:2021yld,Heidenreich:2021xpr,Koide:2021zxj,Kaidi:2021xfk,Choi:2021kmx,Roumpedakis:2022aik,Bhardwaj:2022yxj,Cordova:2022ieu,Bashmakov:2022jtl,Choi:2022rfe,Bartsch:2022mpm,Apruzzi:2022rei,GarciaEtxebarria:2022vzq,Niro:2022ctq,Chen:2022cyw,Bashmakov:2022uek,Karasik:2022kkq,GarciaEtxebarria:2022jky,Choi:2022fgx,Yokokura:2022alv,Bhardwaj:2022maz,Bartsch:2022ytj,Kaidi:2023maf,Lin:2023uvm,Chen:2023czk}, as well as higher-form symmetry mentioned above.    
%
%

Considering such a generalization of particle and symmetry, it is natural to investigate a possibility of constructing a higher-form version of the GL theory. 
In this paper, we propose a field theory of superconducting closed $p$-brane $C_p^{}$ and study its fundamental phenomena at the mean-field (classical) level. 
We should note that the model considered in this paper was first proposed in our previous  paper~\cite{Hidaka:2023gwh} where we have focused only on low-energy effective theory in the superconducting phase.  
%
The purpose of this paper is to present more detailed analysis of the model and show that our brane-field theory is indeed a natural extension of the conventional GL theory to higher-dimensional branes.  

Another important aspect of modern physics is a duality between different theories. 
In the field of superconductivity, a famous one is the Particle-Vortex duality in $3$-dimensional (Minkowski) spacetime~\cite{Peskin:1977kp,Dasgupta:1981zz}. 
It is a duality between the Abelian-Higgs model and complex scalar theory (XY model) with a global $\mathrm{U}(1)$ symmetry near the infra-red (IR) fixed point, i.e. Wilson-Fisher fixed point. 
If we disregard the details, the essence of the duality simply follows from universality argument:  
In a $3$-dimensional spacetime, the dual-form of the gauge field $A_1^{}$ is a real scalar field, which implies that the physical degrees of freedom in the Abelian-Higgs model is the same as a complex scalar field. 
In addition, the Bianchi identity $dF_2^{}=d^2A_1^{}=0$ means the existence of a magnetic $\mathrm{U}(1)$ symmetry, which corresponds to the global $\mathrm{U}(1)$ symmetry in the XY model. 
Thus, both theories have the same degrees of freedom and symmetry and should belong to the same universality class. 

Following our discussion on the construction of the brane-field theories, it is quite natural to expect that a similar duality can hold even among brane-field theories.  
In Section~\ref{sec:duality}, we conjecture a duality between the superconducting brane-field theory and dual-brane field theory with a global $\mathrm{U}(1)$ higher-form symmetry by showing the   correspondence between symmetry, order-parameters, and low-energy excitation modes.   
Although the quantum nature of these theories is not fully understood  yet, we will see that such a duality seems as convincing as the ordinary Particle-Vortex duality from the point of view of universality.   
Since a dual object of brane is no longer particle or vortex in general, we call this new duality {\it Brane-Dual-Brane Duality}.

\

The organization of this paper is as follows. 
In Section~\ref{sec:superconductor}, we introduce the superconducting brane-field model and discuss various fundamental phenomena such as the spontaneous symmetry breaking, Meissner effect, topological defect and so on. 
We explicitly construct a topologically non-trivial static solution (i.e. Vortex solution for $p=0$) and calculate the low-energy effective theory in the presence of it. 
In Section~\ref{sec:duality}, we discuss a duality between the superconducting brane-field theory and dual brane-field theory with a global $\mathrm{U}(1)$ higher-form symmetry. 
%
%
Section~\ref{Summary and Discussion} is devoted to summary and discussion.

\section{Superconducting Brane Field Theory}
\label{sec:superconductor}
We introduce a field theory of a closed $p$-brane $C_p^{}$ interacting with a $(p+1)$-form gauge field $A_{p+1}^{}$, which was  originally proposed in Ref.~\cite{Hidaka:2023gwh}. 
In the following, we represent a $d$-dimensional spacetime with a metric $g_{\mu\nu}^{}(X)$ by $\Sigma_{d}^{}$ and employ the Minkowski metric signature, $(-,+,+,\cdots,+)$.  
A $p$-dimensional spacelike closed brane is represented by $C_p^{}$, which is  expressed by an embedding function $S^p\to \Sigma_d$, i.e. $\{X^\mu(\xi)\}_{\mu=0}^{d-1}$~, where $S^p$ is a $p$-dimensional sphere and $\xi=\{\xi^i\}_{i=1}^p$ denotes intrinsic coordinates of $S^p$. 
Besides, we represent the determinant of the induced metric as 
\aln{
h={\rm det}(h_{ij}^{})~,\quad h_{ij}^{}=\frac{\partial X^\mu(\xi)}{\partial \xi^i}\frac{\partial X^\nu(\xi)}{\partial \xi^j}g_{\mu\nu}^{}(X(\xi))~.
} 
In this notation, the volume of $C_p^{}$ is given by
\aln{{\rm Vol}[C_p^{}]=\int d^p\xi \sqrt{h}~.
}   
Note also that when $\{X^\mu(\xi)\}_{\mu=0}^{d-1}$ represents an embedding of a closed subspace $C_p^{}$, its translation $\{X^\mu(\xi)+x^\mu\}_{\mu=0}^{d-1}$ also represents another closed brane (with a same volume).  

Then, we introduce a complex scalar field $\psi[C_p^{}]$ which is a functional of $\{X^\mu(\xi)\}_{\mu=0}^{d-1}$ and supposed to be a scalar field under the spacetime diffeomorphism and reparametrization on $C_p^{}$.
See Ref.~\cite{Hidaka:2023gwh} for more details.   
%
%

\subsection{Model, Symmetries, and Phases}
We consider the gauged $p$-form brane-field model~\cite{Hidaka:2023gwh}:
\aln{
S={\cal N}\int [dC_{p}^{}]\left\{-
\int_{\Sigma_d^{}}\frac{\delta(C_{p}^{})}{\mathrm{Vol}[C_{p}^{}]}D_{p+1}^{G}\psi^\dagger[C_p^{}]\wedge \star D_{p+1}^{G}\psi^{}[C_p^{}]-V(\psi^\dagger \psi)
\right\}-\frac{1}{2g^2}\int F_{p+2}^{}\wedge \star F_{p+2}^{}~,
\label{toy model}
}
where 
\aln{
&D_{p+1}^{G}\psi[C_p^{}]=D_{p+1}\psi[C_p^{}]-iqA_{p+1}^{}\psi[C_p^{}]~,\quad F_{p+2}^{}=dA_{p+1}^{}~,
\\
&\delta(C_p^{})\coloneqq\int_{C_p^{}}^{} d^p\xi\sqrt{-\frac{h}{g}}\delta^{(d)}\left(X^\mu-X^\mu(\xi)\right)~,
}
$q\in \mathbb{Z}$, $g^2$ is a gauge coupling, $[dC_p^{}]$ is an appropriate path-integral measure of $C_p^{}$\footnote{See Refs.~\cite{Hidaka:2023gwh,Iqbal:2021rkn} for more details about this measure. 
Such details are irrelevant in the following discussion.  
}, and ${\cal N}$ is a normalization factor.  
Here, $A_{p+1}$ is a $\mathrm{U}(1)$ $(p+1)$-form gauge field and 
\aln{
D_{p+1}^{}\psi[C_p^{}]=\frac{1}{(p+1)!}\frac{\delta \psi[C_p^{}]}{\delta \sigma^{\mu_1^{}\cdots \mu_{p+1}^{}}(\xi)}dX^{\mu_1^{}}\wedge \cdots \wedge dX^{\mu_{p+1}^{}}~ 
}
is the area derivative introduced in Ref.~\cite{Hidaka:2023gwh}. 
It is a natural generalization of the ordinary derivative $d\phi(X)=\partial_\mu^{}\phi(X)dX^\mu$ to a higher-dimensional object, and one of the important properties is  
\aln{D_{p+1}^{}\left(\int_{C_p^{}}f_p^{}\right)=df_p^{}
}
for $^\forall $ $p$-form $f_p^{}$. 
%

The action is invariant under the $p$-form gauge transformation:
\begin{equation}
\psi[C_p^{}]\rightarrow e^{iq\int_{C_p^{}}\Lambda_p^{}}
\psi[C_p^{}]~,\quad 
A_{p+1}^{}  \rightarrow  A_{p+1}^{}+d\Lambda_{p}^{}~,
\label{eq:gauge transformation Bp}
\end{equation}
where $\Lambda_p$ is an arbitrary $p$-form normalized as $\int_{C_{p+1}^{}}d\Lambda_{p}^{}\in 2\pi\mathbb{Z}$.

The equation of motion (EOM) of $A_{p+1}^{}$ is 
\aln{\frac{(-1)^p}{g^2}d\star F_{p+2}^{}=q\star J_{p+1}^{}~,
\label{EOM}
}
where 
\aln{
J_{p+1}^{}(X)={\cal N}\int [dC_p^{}]\frac{\delta(C_p^{})}{{\rm Vol}[C_p^{}]}i\bigg\{\psi^\dagger[C_p^{}] D_{p+1}^G\psi[C_p^{}]-\psi[C_p^{}] (D_{p+1}^G\psi[C_p^{}])^\dagger\bigg\}~
\label{p+1 current}
}
is the $(p+1)$-form current of the brane-field.  
Equation~(\ref{EOM}) implies that it is on-shell conserved,
\aln{d\star J_{p+1}^{}=0~,
}
and the conserved charge is 
\aln{
Q_{p+1}^{}=\int_{D_{d-p-1}}^{}\star J_{p+1}^{}=\frac{1}{qg^2}\int_{C_{d-p-2}}^{}\star  F_{p+2}^{}~,
}
where $C_{d-p-2}^{}$ is a $(d-p-2)$-dimensional (spatial) closed subspace and $D_{d-p-1}^{}$ is a $(d-p-1)$-dimensional (spatial) open subspace with the boundary  $\partial D_{d-p-1}^{}=C_{d-p-2}^{}$.  
%
When $q>1$, the above conserved charge implies the existence of a global electric $\mathbb{Z}_q^{}$ $(p+1)$-form symmetry, whose transformation is given by  
\aln{
\label{p+1 form symmetry}
A_{p+1}^{} \rightarrow A_{p+1}^{}+\frac{1}{q}\Lambda_{p+1}^{}~,\quad d\Lambda_{p+1}^{}=0~,\quad  \int_{C_{p+1}^{}}\Lambda_{p+1}^{}\in 2\pi\mathbb{Z}~, 
}
where $C_{p+1}^{}$ is a $(p+1)$-dimensional closed subspace.  
The corresponding symmetry operator and charged operator are
\aln{
U_n^{}[C_{d-p-2}] &=\exp\left(2\pi i nQ_{p+1}^{}\right)= \exp\left(i\frac{2\pi n}{qe^2}\int_{C_{d-p-2}}\star F_{p+2}\right)~,\\
W[C_{p+1}^{}]&=\exp\left(i\int_{C_{p+1}^{}}A_{p+1}^{}\right)~,
}
respectively. 
In addition, the theory has a magnetic $\mathrm{U}(1)$ $(d-p-3)$-form symmetry
\aln{dF_{p+2}^{}=0~,
} 
whose charge is 
\aln{Q_{d-p-3}=\frac{1}{2\pi}\int_{C_{p+2}}F_{p+2}\in\mathbb{Z}~,
}
 and the charged object is $(d-p-3)$-dimensional 't Hooft operator
\aln{T[C_{d-p-3}^{}]:=\exp\left(i\int_{C_{d-p-3}^{}}\tilde{A}_{d-p-3}^{}\right)~,
 }
 where $C_{d-p-3}^{}$ is a $(d-p-3)$ dimensional closed subspace and $\tilde{A}_{d-p-3}^{}$ is the dual field of $A_{p+1}^{}$. 
In the following, we represent this symmetry by $\mathrm{U}_{\rm M}^{}(1)$.

\

\noindent {\bf COULOMB PHASE}\\
When the $p$-form gauge symmetry (\ref{eq:gauge transformation Bp}) is not broken, i.e. $\langle \psi[C_p^{}]\rangle=0$, the theory has massive brane-field modes and massless $(p+1)$-form $A_{p+1}^{}$, where the latter corresponds to the Nambu-Goldstone modes of the broken magnetic $(d-p-3)$-form symmetry $\mathrm{U}_M^{}(1)$.    
Thus, the low-energy effective theory is simply the pure $(p+1)$-form Maxwell  theory
\aln{S_{\rm eff}^{}=-\frac{1}{2g^2}\int F_{p+2}^{}\wedge \star F_{p+2}^{}~,\quad F_{p+2}^{}=dA_{p+1}^{}~,
}
which has an emergent electric $\mathrm{U}(1)$ $(p+1)$-form symmetry given by
\aln{
A_{p+1}^{}~\rightarrow ~A_{p+1}^{}+\Lambda_{p+1}~,\quad d\Lambda_{p+1}^{}=0~,\quad \int_{C_{p+1}^{}}\Lambda_{p+1}^{}\in 2\pi \mathbb{Z}~. 
} 
This symmetry is also spontaneously broken for $d>p+2$~\cite{Lake:2018dqm}.

\

\noindent {\bf SUPERCONDUCTING PHASE}\\
When the $p$-form gauge symmetry (\ref{eq:gauge transformation Bp}) is broken, i.e. $\langle \psi[C_p^{}]\rangle=v\neq 0$, the gauge field $A_{p+1}^{}$ becomes massive and the low-energy effective theory is described by a topological field theory~\cite{Hidaka:2023gwh}.  
 In particular, the effective theory has an emergent $(d-p-2)$-form symmetry, which can be intuitively understood as follows.    
 In the vacuum, $\langle \psi[C_p^{}]\rangle=v~,\langle A_{p+1}^{}\rangle=0$,  the relevant fluctuation is a phase modulation of the brane-field defined by 
 \aln{\psi[C_p^{}]=v\exp\left(i\int_{C_p^{}}B_p^{}\right)~,
 } 
 and the $(p+1)$-form current (\ref{p+1 current}) becomes  
 \aln{J_{p+1}^{}\sim v^2dB_{p+1}^{}~,
 } 
 which satisfies $dJ_{p+1}^{}=0$.
This can be understood as a conservation law for a $(d-p-2)$-dimensional object. 
In the present model (\ref{toy model}), such a $(d-p-2)$-dimensional object is described by a topologically non-trivial solution in the superconducting phase, as we will 
 %
see in Subsection~\ref{sec:soliton}.  

Note also that the magnetic $(d-p-3)$-form symmetry $\mathrm{U}_M^{}(1)$ is not broken in the superconducting phase because of the confinement of the $p$-form magnetic field, i.e. the Meissner effect, which is explicitly discussed in the following section.  
%
Correspondingly, the expectation value of the 't Hooft operator in Euclidean spacetime indicates the area law\footnote{This is shown as follows. 
For a given time slice, the exponent of the 't Hooft operator corresponds to the energy between two magnetically charged $(d-p-4)$-branes $C^{\rm M}_{d-p-4}$. 
Because of the confinement of the magnetic flux, the energy per unit volume (i.e. tension) of $C^{\rm M}_{d-p-4}$ is linearly dependent on the distance $R$ between two branes. 
Thus, in the large volume limit,
\aln{
\langle T[C_{d-p-3}^{}]\rangle\sim \exp\left(-c\times TR\times {\rm Vol}[C^{\rm M}_{d-p-4}]\right)=\exp\left(-c\times {\rm Vol}[M_{d-p-2}^{}]\right)~,
}
where $T$ is a stretched length of $C_{d-p-3}^{}$ in the time direction. 
} 
\aln{\langle T[C_{d-p-3}^{}]\rangle \sim \exp\left(-c\times {\rm Vol}[M_{d-p-2}^{}]\right)\quad \text{for }{\rm Vol}[C_{d-p-3}^{}]\rightarrow \infty~,
} 
where $c$ is a positive constant and $M_{d-p-2}^{}$ is the $(d-p-2)$-dimensional minimal surface enclosed by $C_{d-p-3}^{}$.

\subsection{Meissner Effect}
Here we discuss a higher-form version of the Meissner effect.  
We assume that the brane-field potential $V(\psi^\dagger\psi)$ has a nontrivial minimum $\langle\psi[C_p^{}]\rangle=v\neq 0$.  
In the vacuum, the current~(\ref{p+1 current}) is evaluated as 
\aln{
J_{p+1}^{}(X)&=2qv^2\times {\cal N}\int [dC_p^{}]\frac{1}{{\rm Vol}[C_p^{}]}\int d^p\xi\sqrt{h}A_{p+1}^{}(X(\xi))\delta^{(d)}(X-X(\xi))
\nn
&=2qv^2A_{p+1}(X)\times {\cal N}\int [dC_p^{}]\frac{1}{{\rm Vol}[C_p^{}]}\int d^p\xi\sqrt{h}\delta^{(d)}(X-X(\xi))~,
}
where the remaining path-integral does not depend on the coordinate $X^\mu$ because $\int [dC_p^{}]$ contains the zero-mode integrations, $\int d^dx$.  
Thus we can take the normalization factor ${\cal N}$ so that 
\aln{
J_{p+1}^{}(X)=2qv^2(-1)^{p(d-p-1)}A_{p+1}^{}(X)~,
}
which corresponds the London equation for $p=0$.  
Now the EOM (\ref{EOM}) becomes 
\aln{
\frac{1}{g^2}\star d\star F_{p+2}^{}=-2(qv)^2A_{p+1}^{}~. 
}  
In the Minkowski spacetime, this is explicitly written as 
\aln{
\partial_\mu^{}({F^\mu}_{\mu_1^{}\cdots \mu_{p+1}^{}})=2(qvg)^2(A_{p+1})_{\mu_1^{}\cdots \mu_{p+1}^{}}~. 
\label{EOM in component}
}
By taking the Lorentz gauge 
\aln{
\partial^\mu (A_{p+1}^{})_{\mu \mu_{2}^{}\cdots \mu_{p+1}^{}}=0\quad \text{for}~^\forall~\mu_2^{},\cdots,\mu_{p+1}^{}~,
} 
Eq.~(\ref{EOM in component}) becomes
\aln{
\Box (A_{p+1}^{})_{\mu_1^{}\cdots \mu_{p+1}^{}}^{}=2(qvg)^2(A_{p+1})_{\mu_1^{}\cdots \mu_{p+1}^{}}~,
\label{Meissner}
}
which implies that the mass of $A_p^{}$ is $m_A^{}=\sqrt{2}qvg$. 
When the gauge field is static, Eq.~(\ref{Meissner}) means that the penetration depth of the $p$-form magnetic field is  
\aln{
\lambda=\frac{1}{m_A^{}}~. 
}

\subsection{Topological Defect}\label{sec:soliton}

\begin{figure}
    \centering
    \includegraphics[scale=0.15]{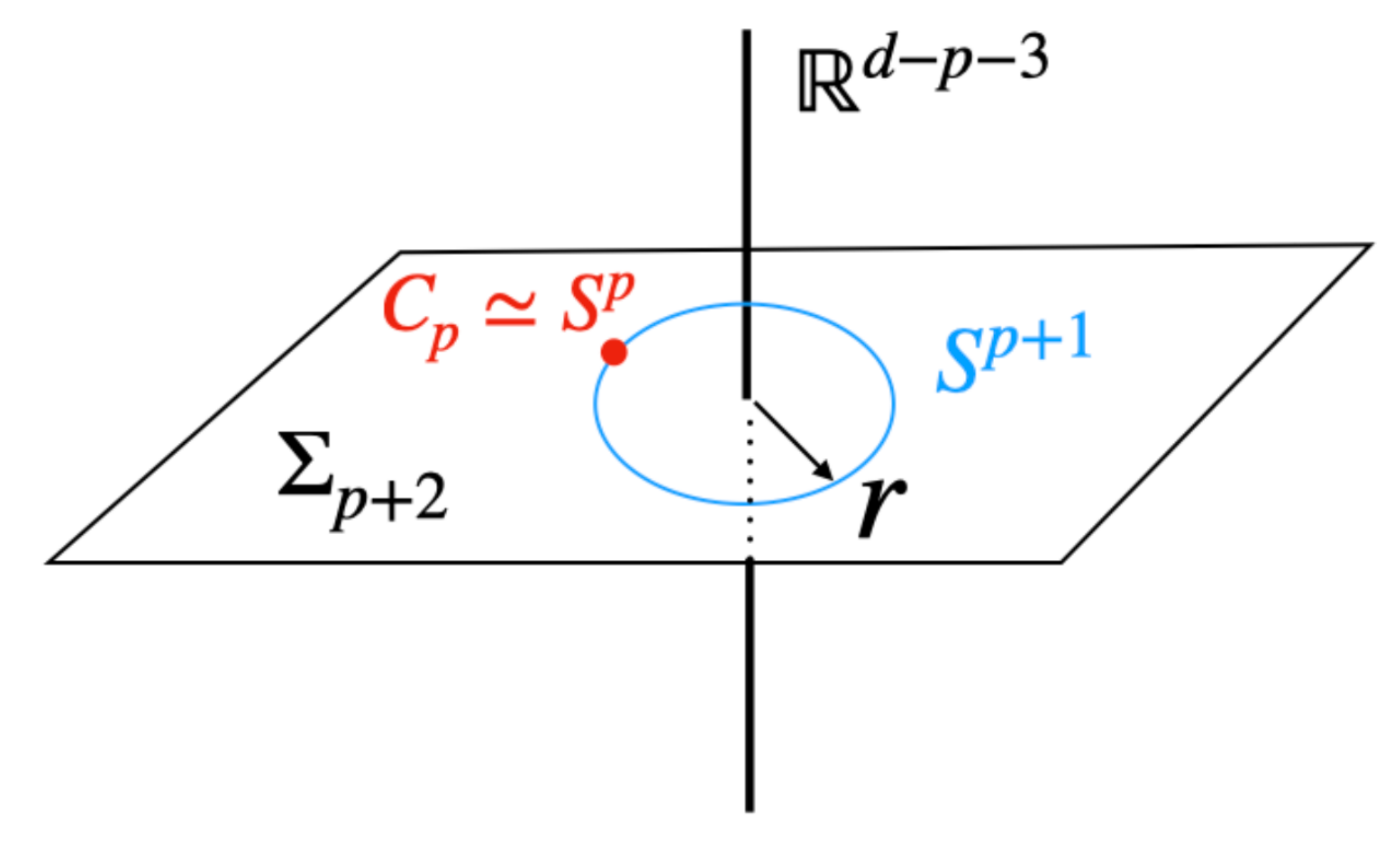}
     \includegraphics[scale=0.15]{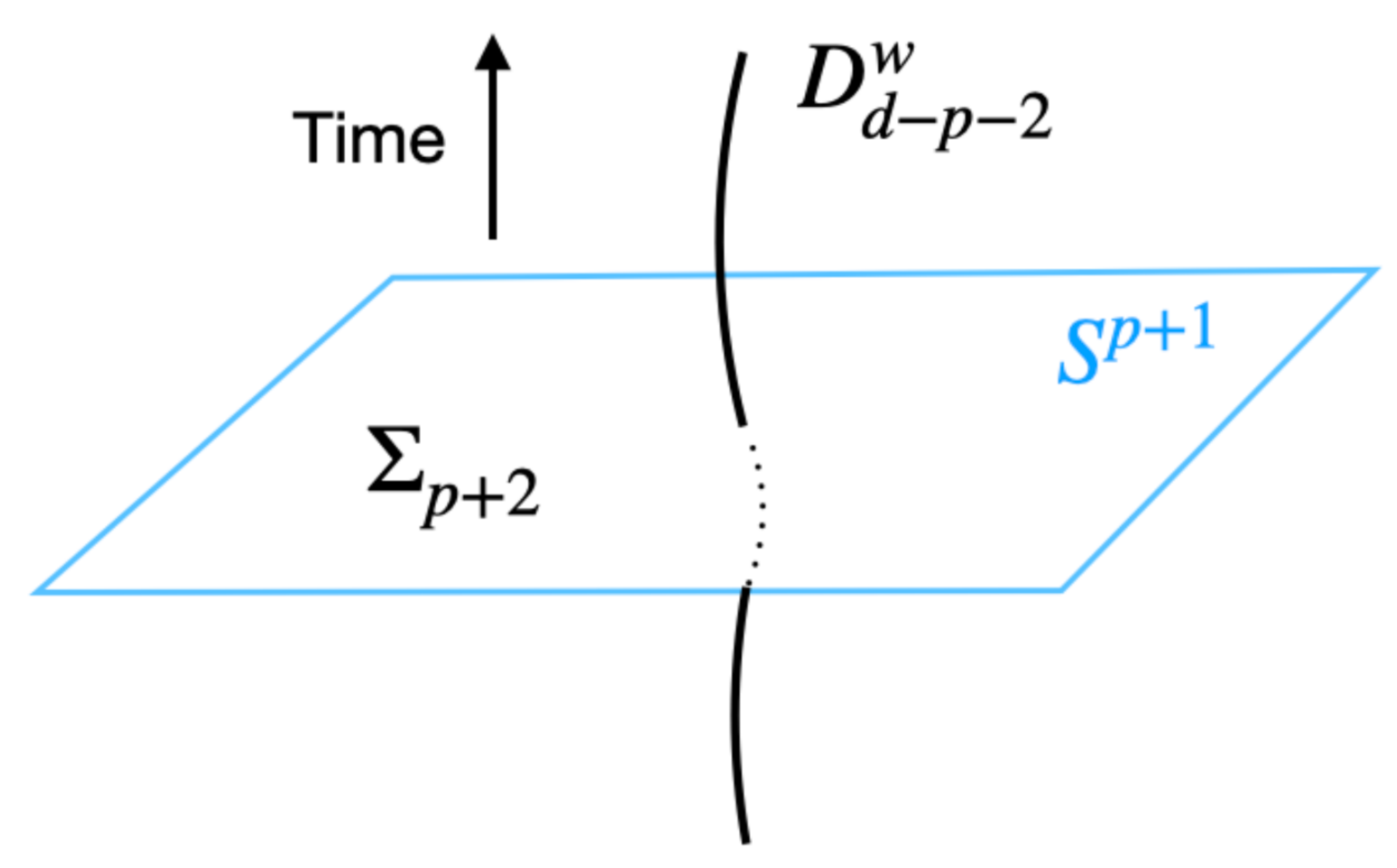}
    \caption{Left: Topologically non-trivial static configuration of the brane-field theory (\ref{toy model}). 
    The red point corresponds to $p$-brane $C_p^{}\simeq  S^p$. 
    Right: The linking between the world-volume $D^W_{d-p-2}$ and $S^{p+1}$. 
    }
    \label{fig:soliton}
\end{figure}

As in the ordinary Ginzburg-Landau theory, the brane-field theory (\ref{toy model}) allows a topologically non-trivial static configuration in the superconducting phase. 
For simplicity, we consider the flat spacetime $\Sigma_d^{}=\mathbb{R}\times \mathbb{R}^{d-1}$ and represent the space as 
\aln{
\mathbb{R}^{d-1}=\mathbb{R}^{d-p-3}\times S^{p+1}\times [0,\infty)~,
}
where $r\in [0,\infty)$ denotes the radius of the $(p+1)$-dimensional sphere $S^{p+1}$. 
See the left panel in Fig.~\ref{fig:soliton} for example. 
In particular, we represent the $(p+2)$-dimensional subspace as 
\aln{\Sigma_{p+2}^{}:=S^{p+1}\times [0,\infty)~,
} 
which has the boundary $\partial \Sigma_{p+2}^{}= \Sigma_{p+2}^{}|_{r=\infty}^{}=S^{p+1}$. 
Besides, we represent the volume-form of the $q$-dimensional sphere  $S^q$ as $\Omega_q^{}$, i.e. 
\aln{
\int_{S^q} \Omega_q^{}={\rm Area}(S^{q})~.
}
For $q=0$, we define Area$(S^0)=1$. 

We consider $p$-brane $C_p^{}\simeq S^p$ embedded in $S^{p+1}$ specified by $r=$constant and $\theta_1^{}=$ constant where $\theta_1^{}$ is one of the polar coordinates $\theta_1^{}\in [0,\pi]$ of $S^{p+1}$.  
This is shown by  a red point in Fig.~\ref{fig:soliton}. 
This treatment corresponds to restricting the embeddings of $C_p^{}$ as 
\aln{
{\cal N}\int [dC_p^{}]\quad \rightarrow \quad {\cal N}\begin{cases}\int_0^\infty dr r^{p+1}\int_0^{\pi} d\theta_1^{}\sin^{p} (\theta_1^{}) & \text{for }p\geq 1
\\
\int_0^\infty dr r^{}\int_0^{2\pi} d\theta_1^{}  & \text{for }p=0
\end{cases}
~,
\label{minispace}
}  
and the volume of $C_p^{}$ is given by
\aln{
{\rm Vol}[C_p^{}]=(r\sin\theta_1^{})^p\int_{S^{p}}\Omega_p^{}
=(r\sin\theta_1^{})^p\times \mathrm{Area}(S^{p})~.
}
%
%
We then consider a following ansatz: 
\aln{
\psi[C_p^{}]=\left(\int_{S^{p}}\chi_p^{}\right)
\exp\left(i\int_{S_p^{}}B_{p}^{W}\right)~,\quad A_{p+1}^{}=A_{}^{}(r)\Omega_{p+1}^{W}~.
\label{static configuration}
}
with 
\aln{
\chi_p^{}
=\chi(r)(r\sin\theta_1^{})^p\Omega_p^{}~,\quad \chi(r)\in \mathbb{R}~,
}
and $\Omega_{p+1}^W:=dB_{p}^{W}$ is the volume-form of $S^{p+1}$ normalized as\footnote{Explicitly, it is given by 
\aln{
dB_p^{W}&=N\sin^{p}(\theta_1^{})\sin^{p-1}(\theta_2^{})\cdots \sin(\theta_p^{})d\theta_1^{}\wedge d\theta_2^{}\wedge \cdots \wedge d\theta_{p+1}^{}
\\
&=\frac{N}{p+1}d\left[\sin^{p+1}(\theta_1^{})\sin^{p-1}(\theta_2^{})\cdots \sin(\theta_p^{})d\theta_2^{}\wedge \cdots \wedge d\theta_{p+1}^{}\right]~,
} 
where $N$ is a normalization factor. 
}
\aln{
\int_{S^{p+1}} \Omega_{p+1}^{W}=2\pi n~,\quad n\in \mathbb{Z}~,
\label{theta normalization}
}   
which guarantees the single-valuedness of the brane-field in the volumeless limit\footnote{For $\theta_1^{}\rightarrow 0,\pi$, the volume of $C_p^{}$ becomes zero, and we demand that the brane-field $\psi[C_p^{}]$ is single valued for such a volumeless limit. 
In principle, this requirement should be justified by microscopic (UV) physics.
}
\aln{\psi[C_p^{}|_{\theta_1^{}=0}]=\psi[C_p^{}|_{\theta_1^{}=\pi}]~.
}
Thus, the ansatz (\ref{static configuration}) is characterized by the first homotopy group $\pi_1^{}(\mathrm{U}(1)\simeq S^1)$.

Denoting the world-volume of the defect as 
\aln{
D^W_{d-p-2}=\mathbb{R}\times \mathbb{R}^{d-p-3}~,
}
we can also relate $\Omega_{p+1}^{W}$ to the world-volume as 
\aln{\delta_{p+2}^{}(D^W_{d-p-2})=\frac{1}{2\pi n}d(f(r)\Omega_{p+1}^W)~,
}
where $f(r)$ is a smooth function satisfying $f(r=\infty)=1$, and 
 $\delta_{p+2}^{}(D^W_{d-p-2})$ is the Poincare-dual form defined by
\aln{
\int_{D^W_{d-p-2}} f_{d-p-2}^{}=\int_{\Sigma_d^{}}f_{d-p-2}^{}\wedge \delta_{p+2}^{}(D^W_{d-p-2})~,
}
for $^\forall~(d-p-2)$-form $f_{d-p-2}^{}$. 
In particular, we have
\aln{
\int_{\Sigma_{d}^{}}\delta_{p+2}^{}(D^W_{d-p-2})\wedge \delta_{d-p-2}^{}(\Sigma_{p+2}^{})=\int_{\Sigma_{p+2}}\delta_{p+2}^{}(D^W_{d-p-2})=\frac{1}{2\pi n}\int_{\Sigma_{p+2}}d\Omega^W_{p+1}=\frac{1}{2\pi n}\int_{S^{p+1}}\Omega_p^W=1~,
}
which corresponds to the linking between $D_{d-p-2}^W$ and $S_{}^{p+1}$. 
See the right panel in Fig.~\ref{fig:soliton} for example. 

The exterior derivative of $\chi_p^{}$ is evaluated as 
\aln{d\chi_p^{}&=\sin^p\theta_1^{}\frac{d(r^p\chi(r))}{dr}dr\wedge \Omega_p^{}+pr^p\chi(r)\sin^{p-1}\theta_1^{}d\theta_1^{}\wedge \Omega_p^{}
\\
&=\sin^p\theta_1^{}\frac{d(r^p\chi(r))}{dr}dr\wedge \Omega_p^{}+\frac{pr^p\chi(r)}{\sin\theta_1^{}}\frac{{\rm Area}(S^{p+1})}{2\pi n}\Omega^W_{p+1}~.
}
%
The kinetic term of the brane-field is now evaluated as 
\aln{
&{\cal L}_{\rm Kin}:=\frac{1}{\mathrm{Vol}[C_p^{}]}\int_{\Sigma_d^{}}\delta(C_p^{}) \bigg(d\chi_p^{}-i\chi(r){\rm Vol}[C_p^{}](1-qA_{}^{})\Omega_{p+1}^W
\bigg)\wedge \star \bigg(d\chi_p^{}+i\chi(r){\rm Vol}[C_p^{}](1-qA_{}^{})\Omega_{p+1}^W
\bigg)
\nn
=&\frac{1}{\mathrm{Vol}[C_p^{}]}\int_{\Sigma_d^{}}d^dX\delta(C_p^{})\sqrt{-g}\bigg[\frac{1}{r^{2p}}\left(\frac{d(r^p\chi)}{dr}\right)^2+\frac{1}{r^{2(p+1)}}\left(\frac{2\pi n r^p\chi}{{\rm Area}(S^{p+1})}\right)^2\bigg\{\left(\frac{p}{\sin\theta_1^{}}\frac{{\rm Area}(S^{p+1})}{2\pi n}\right)^2
\nn
&\hspace{4cm}+\bigg(\sin^p \theta_1^{}{\rm Area}(S^p)(1-qA(r))\bigg)^2\bigg\}
\bigg]
\nn
=&\left[\frac{1}{r^{2p}}\left(\frac{d(r^p\chi)}{dr}\right)^2+\frac{1}{r^{2(p+1)}}\left(\frac{2\pi n r^p\chi}{{\rm Area}(S^{p+1})}\right)^2\bigg\{\left(\frac{p}{\sin\theta_1^{}}\frac{{\rm Area}(S^{p+1})}{2\pi n}\right)^2+\bigg(\sin^p \theta_1^{}{\rm Area}(S^p)(1-qA(r))\bigg)^2
\right].
} 
By using Eq.~(\ref{minispace}), the effective action is 
\aln{S=-{\cal N}\int_{\mathbb{R}\times \mathbb{R}^{d-p-3}}&dtd^{d-p-3}X\int_0^\infty drr^{p+1}\int_0^\pi d\theta_1^{}\sin^p\theta_1^{}
\nn
&\times \bigg[{\cal L}_{\rm Kin}+V\left((\chi{\rm Vol}[C_p^{}])^2\right)
+\frac{
1}{2g^2r^{2(p+1)}}\left(\frac{2\pi n}{{\rm Area}(S^{p+1})}\right)^2\left(\frac{dA}{dr}\right)^2
\bigg]~,
}
which leads to the energy density as\footnote{In Eq.~(\ref{general soliton action}), the quadratic term of $f(r)$ is divergent for $p=1$ due to $\Gamma\left(\frac{p-1}{2}\right)$.  
This term originally comes from the integration $\int_{0}^\pi d\theta_1^{}\sin^{p-2}\theta_1^{}$, which is apparently divergent for $p=1$ and $\theta_1^{}\rightarrow 0,\pi$. 
Since $\theta_1^{}\rightarrow 0,\pi$ correspond to the volumeless limit of $C_p^{}$, this is an UV divergence and should be treated carefully by microscopic physics.
In the numerical calculations below, we simply neglect the divergent term for $p=1$.   
We should also note that the existence of such a divergence depends on how to choose the path integral measure $[dC_p^{}]$ as well.  
} 
\aln{
&E_{d-p-3}^{}=\int_0^\infty drr^{p+1}\int_0^\pi d\theta_1^{}\sin^p\theta_1^{}\bigg[{\cal L}_{\rm Kin}
+V\left((\chi{\rm Vol}[C_p^{}])^2\right)
+\frac{
1}{2g^2r^{2(p+1)}}\left(\frac{2\pi n}{{\rm Area}(S^{p+1})}\right)^2\left(\frac{dA}{dr}\right)^2
\bigg]
\\
=&\frac{\sqrt{\pi}\Gamma(\frac{p+1}{2})}{\Gamma(\frac{p+2}{2})}\int_0^\infty drr^{p+1}\bigg[\frac{1}{r^{2p}}\left(\frac{df}{dr}\right)^2
+\frac{f^2}{r^{2(p+1)}}\frac{\Gamma(\frac{p+2}{2})}{\Gamma(\frac{p+1}{2})}\bigg\{\frac{\Gamma(\frac{p-1}{2})}{\Gamma(\frac{p}{2})}p^2
+\frac{\Gamma(\frac{3p+1}{2})}{\Gamma(\frac{3p+2}{2})}\bigg(\frac{2\pi n{\rm Area}(S^p)}{{\rm Area}(S^{p+1})}(1-qA(r))\bigg)^2\bigg\}
\nn
&\hspace{3cm}+
V(f)+\frac{
1}{2g^2}\left(\frac{2\pi n}{{\rm Area}(S^{p+1})}\right)^2\frac{1}{r^{2(p+1)}}\left(\frac{dA}{dr}\right)^2
\bigg]~,
\label{general soliton action}
}
where $f(r)=r^p\chi(r)$. 
%
In particular, for $p=1$ and 
\aln{V(f)=\frac{\lambda}{4}(f^2-v^2)^2~,
\label{wine-bottle}
}
Eq.~(\ref{general soliton action}) corresponds to the tension of the Abrikosov-Nielsen-Olesen vortex~\cite{Abrikosov:1957wnz,Nielsen:1973cs,Eto:2022hyt}
\aln{E_{d-3}^{}=2\pi\int_0^{\infty}drr\left[\left(\frac{df}{dr}\right)^2+\frac{n^2}{r^2}f^2(1-qA(r))^2+\frac{\lambda}{4}(f^2-v^2)^2+\frac{n^2}{2g^2r^2}\left(\frac{dA}{dr}\right)^2
\right]~.
} 
One can see that $f(r)$ and $A(r)$ must satisfy the boundary conditions 
\aln{f(r)\rightarrow v~,\quad qA(r)\rightarrow 1\quad \text{for $r\rightarrow \infty$}~
}
in order that the configuration has a finite-energy density for general $p$.
From Eqs.~(\ref{static configuration})(\ref{theta normalization}) and the above boundary condition, we obtain the quantization of the magnetic flux as 
\aln{\int_{\Sigma_{p+2}^{}}F_{p+2}^{}=\int_{S^{p+1}}qA_{p+1}^{}=2\pi n\in 2\pi \mathbb{Z}~, 
}
where we have used the Stokes theorem in the first equality.   

To study the EOMs of Eq.~(\ref{general soliton action}), it is convenient to introduce dimensionless quantities by 
\aln{f=v\times \overline{f}~,\quad r=\frac{y}{gv}~,
} 
and Eq.~(\ref{general soliton action}) is rewritten as 
\aln{
\frac{E_{d-p-3}^{}}{v^2(gv)^p}&\propto \int_0^\infty dy y^{p+1}\bigg[\frac{1}{y^{2p}}\left(\frac{d\overline{f}}{dy}\right)^2
+\frac{\overline{f}^2}{y^{2(p+1)}}\frac{\Gamma(\frac{p+2}{2})}{\Gamma(\frac{p+1}{2})}\bigg\{\frac{\Gamma(\frac{p-1}{2})}{\Gamma(\frac{p}{2})}p^2
+\frac{\Gamma(\frac{3p+1}{2})}{\Gamma(\frac{3p+2}{2})}\bigg(\frac{2\pi n{\rm Area}(S^p)}{{\rm Area}(S^{p+1})}(1-qA(r))\bigg)^2\bigg\}
\nn
&\hspace{3cm}+
\frac{1}{(gv)^{2p+2}v^{2}}V(\overline{f})+\frac{
1}{2}\left(\frac{2\pi n}{{\rm Area}(S^{p+1})}\right)^2\frac{1}{y^{2(p+1)}}\left(\frac{dA}{dy}\right)^2
\bigg]~.
}
By taking the variation of this equation with respect to $\overline{f}$ and $A$, we obtain  
\aln{
&\frac{1}{y^{p+1}}\frac{d}{dy}\left(\frac{1}{y^{p-1}}\frac{d\overline{f}}{dy}\right)-\frac{\overline{f}}{y^{2(p+1)}}\frac{\Gamma(\frac{p+2}{2})}{\Gamma(\frac{p+1}{2})}\bigg\{\frac{\Gamma(\frac{p-1}{2})}{\Gamma(\frac{p}{2})}p^2
+\frac{\Gamma(\frac{3p+1}{2})}{\Gamma(\frac{3p+2}{2})}\bigg(\frac{2\pi n{\rm Area}(S^p)}{{\rm Area}(S^{p+1})}(1-qA(r))\bigg)^2\bigg\}
\nn
&\hspace{4cm}-\frac{1}{2(gv)^{2p+2}v^2}\frac{\partial V(\overline{f})}{\partial \overline{f}}=0~,
\label{EOM1}
\\
&\frac{d}{dy}\left(\frac{1}{y^{p+1}}\frac{dA}{dy}\right)+2q{\rm Area}(S^p)^2\frac{\overline{f}^2}{y^{p+1}}\frac{\Gamma(\frac{p+2}{2})\Gamma(\frac{3p+1}{2})}{\Gamma (\frac{p+1}{2})\Gamma (\frac{3p+2}{2})}(1-qA)=0~.
\label{EOM2}
}
In the case of the wine-bottle potential (\ref{wine-bottle}), its dimensionless form is 
\aln{\frac{1}{(gv)^{2p+2}v^2}V(f)=\frac{\overline{\lambda}}{4}(\overline{f}^2-1)^2~,\quad \overline{\lambda}=\frac{\lambda}{g^{2p+2}v^{p}}~.
}
\begin{figure}
    \centering
    \includegraphics[scale=0.6]{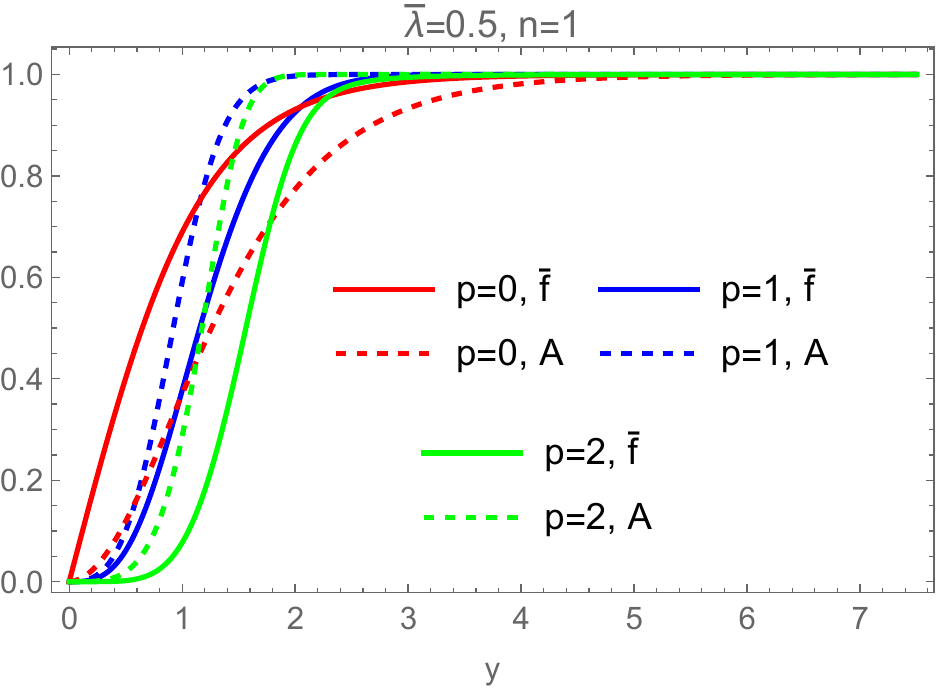}
    \caption{Field profiles of the topological defect for $p=0,1,2$. 
    }
    \label{fig:profile}
\end{figure}
\noindent In Fig.~\ref{fig:profile}, we present the numerical results of the field profiles of the topological defect for $p=0$ (red), $1$ (blue), $2$ (green), where the solid contours correspond to $\overline{f}(y)$ while the dashed contours correspond to $A(y)$. 
Here, the dimensionless coupling $\overline{\lambda}$ is fixed to be $0.5$ and $n=1$.  
%
One can see that the slopes of the fields around the origin are decreasing with increasing $p$. 
This is due to the different $r$ dependence in the kinetic terms, i.e. first terms in Eqs.~(\ref{EOM1})(\ref{EOM2}).  

\subsection{Low energy effective theory}

Let us study the low-energy effective theory in the superconducting phase in the presence of the topological defect.    
We focus on a phase fluctuation of the brane field:
\aln{
\psi[C_{p}^{}]=v\exp\left(
i\int_{C_p^{}} \left(B_p^{W}+B_{p}^{}\right)\right)~.
}
By putting this into Eq.~(\ref{toy model}), we obtain~\cite{Hidaka:2023gwh}
\aln{
-\int_{\Sigma_d^{}}&\left[\frac{1}{2g^2}F_{p+2}^{}\wedge \star F_{p+2}^{}+\frac{\Lambda}{2(2\pi)}(\Omega_{p+1}^W+H_{p+1}^{}-q_{}^{}A_{p+1}^{})\wedge \star (\Omega_{p+1}^W+H_{p+1}^{}-q_{}^{}A_{p+1}^{})\right]~,
\label{effective action}
}
where $\Lambda$ is a parameter whose mass dimension is $d-2(p+1)$, and $H_{p+1}^{}=dB_{p}^{}$.  
In addition to the original $p$-form gauge symmetry~\eqref{eq:gauge transformation Bp}, Eq.~(\ref{effective action}) has a $(p-1)$-form gauge symmetry given by
\aln{
B_p^{}\quad &\rightarrow \quad B_p^{}+d\Lambda_{p-1}^{}~. 
}
%
%
Besides, this effective theory has an emergent $\mathbb{Z}_q^{}$ $(d-p-2)$-form symmetry, whose charge is given by 
\aln{
Q_{d-p-2}=\frac{1}{2\pi}\int_{C_{p+1}^{}}H_{p+1}^{}\in \mathbb{Z}~,
}
where $C_{p+1}^{}$ is a $(p+1)$-dimensional closed subspace in spacetime.   
The charged object is the $(d-p-2)$-dimensional 't Hooft operator, which 
%
can be expressed by using a field in the dualized  theory as follows.

By introducing a dual field of $B_p^{}$ as $\tilde{B}_{d-p-2}^{}$~, Eq.~(\ref{effective action}) can be dualized as~\cite{Hidaka:2023gwh}
\aln{ 
\label{dual theory 0}
&-\int_{\Sigma_d^{}}\left[
\frac{1}{2g^2}F_{p+2}^{}\wedge \star F_{p+2}^{}+\frac{q_{}^{}}{2\pi}(\Omega_{p+1}^W+A_{p+1}^{})\wedge \tilde{H}_{d-p-1}^{}
+\frac{1}{2(2\pi)\Lambda}\tilde{H}_{d-p-1}^{}\wedge \star \tilde{H}_{d-p-1}^{}\right]
\\
=&-\int_{\Sigma_d^{}}\left[
\frac{1}{2g^2}F_{p+2}^{}\wedge \star F_{p+2}^{}+\frac{q_{}^{}}{2\pi}A_{p+1}^{}\wedge \tilde{H}_{d-p-1}^{}
+\frac{1}{2(2\pi)\Lambda}\tilde{H}_{d-p-1}^{}\wedge \star \tilde{H}_{d-p-1}^{}\right]-n q_{}^{}\int_{D^W_{d-p-2}}\tilde{B}_{d-p-2}^{}
}
where $\tilde{H}_{d-p-1}=d\tilde{B}_{d-p-2}$. 
%
%
In the dualized theory, the $\mathbb{Z}_q^{}$ $(d-p-2)$-form symmetry is explicitly given by
\aln{
\tilde{B}_{d-p-2}^{}~\rightarrow~\tilde{B}_{d-p-2}^{}+\frac{1}{q}\tilde{\Lambda}_{d-p-2}^{}~,\quad d \tilde{\Lambda}_{d-p-2}^{}=0~,\quad \int_{{C}_{d-p-2}}\tilde{\Lambda}_{d-p-2}^{}\in2\pi\mathbb{Z}~,
}
where $C_{d-p-2}^{}$ is a $(d-p-2)$-dimensional closed subspace in spacetime. 
The 't Hooft operator is now expressed as 
\aln{
T[C_{d-p-2}^{}]&=\exp\left(i\int_{C_{d-p-2}^{}}\tilde{B}_{d-p-2}^{}\right)~. 
}
One can also see that the dual field $\tilde{B}_{d-p-2}^{}$ couples to the world-volume $D^W_{d-p-2}$ of the topological defect in Eq.~(\ref{dual theory 0}). 
%
In the low-energy limit, we can neglect all the kinetic terms and have
\aln{
S_{\rm eff}^{}=-\frac{q}{2\pi}\int_{\Sigma_d^{}}dA_{p+1}^{}\wedge \tilde{B}_{d-p-2}^{}-nq \int_{D^W_{d-p-2}}\tilde{B}_{d-p-2}^{}~, 
}
where the first term is a BF-type topological field theory, and the system  exhibits topological order~\cite{Hidaka:2023gwh}. 

\section{Duality Conjecture}\label{sec:duality}

We conjecture a duality between the superconducting $p$-brane theory and $(d-p-3)$-brane field theory with a global $\mathrm{U}(1)$ higher-form  symmetry. 
For $d=3$ and $p=0$, this corresponds to the ordinary Particle-Vortex duality. 

\subsection{Particle Vortex Duality}
Particle-Vortex duality~\cite{Peskin:1977kp,Dasgupta:1981zz,Karch:2016sxi,Turner:2019wnh,Seiberg:2016gmd} is an infra-red (IR) duality in a $3$-dimensional (Minkowski) spacetime $\Sigma_3^{}$  between the theory of a complex scalar field $\phi(X)$ (XY model) and the Abelian-Higgs model near the IR fixed point (i.e. Wilson-Fisher fixed point): 
\aln{Z_{\rm XY}^{}[A]\simeq Z_{\rm AH}^{}[A]\quad \text{near the IR fixed point}~,
}
where 
\aln{
Z_{\rm XY}^{}[A]&=\int {\cal D}\phi e^{iS_{\rm XY}^{}}~,
\\
S_{\rm XY}^{}&=-\int_{\Sigma_3^{}}\left[(D_A^{}\phi)^\dagger \wedge \star (D_A^{}\phi)-V(\phi^\dagger \phi)\mathbf{1}\right]~,\quad D_A^{}\phi=d\phi-iA_1^{}\phi~, 
\label{XY model}
}
and 
\aln{Z_{\rm AH}^{}[A]&=\int {\cal D}\Phi{\cal D}a e^{iS_{\rm AH}^{}}~,
\\
S_{\rm AH}^{}&=-\int_{\Sigma_3^{}}\left[(D_a^{}\Phi)^\dagger \wedge \star (D_a^{}\Phi)+\tilde{V}(\Phi^\dagger \Phi)\mathbf{1}+\frac{1}{2e^2}F_2^{}\wedge \star F_2^{}\right]+\frac{1}{2\pi}\int_{D_4^{}}da_1^{}\wedge dA_1^{}~,
\\
&\quad D_a^{}\Phi=d\Phi-ia_1^{}\Phi~, \quad F_2^{}=da_1^{}~. 
} 
Here, $A_1^{}=A_\mu^{}(X)dX^\mu$ is a background field, $a_1^{}=a_\mu^{}(X)dX^\mu$ is a dynamical gauge field, and $D_4^{}$ is a $4$-dimensional open manifold with $\partial D_{4}^{}=\Sigma_3^{}$. 
The XY model~(\ref{XY model}) has a global $\mathrm{U}(1)$ symmetry $\phi\rightarrow e^{i\theta}\phi$, which corresponds to the magnetic $\mathrm{U}(1)$ ($0$-form) symmetry in the Abelian-Higgs model whose order-parameter is the 't Hooft operator (two-point function) $T[C_0^{}]=T(X)$.
%
%

The essence of the duality is as follows. 
In the Coulomb phase of the Abelian Higgs model, the 't Hooft operator $T(X)$ develops a nonzero VEV $\langle T(X)\rangle\neq 0$, which means that  the magnetic $\mathrm{U}(1)$ symmetry is spontaneously broken. 
Thus, this phase corresponds to the broken phase of the dual complex scalar theory~(\ref{XY model}). 
The excitation modes are massive $\Phi$ and massless $a_1^{}$, where the latter is essentially equivalent to a massless scalar field because the dual-form of $a_1^{}$ is a real $0$-form in a $3$-dimensional spacetime.  
In the dual complex scalar theory (\ref{XY model}), such a massless scalar corresponds to the NG boson in the broken phase.  
Besides, the dual complex scalar theory (\ref{XY model}) has a (global) vortex solution in the broken phase, and this corresponds to the massive $\Phi^\dagger$ state in the Abelian-Higgs model.\footnote{The anti-vortex corresponds to the anti-particle created by $\Phi$. 
} 

In the superconducting phase, the 't Hooft operator $T(X)$ indicates the area law $\langle T(X)\rangle\sim e^{-m|X|}$, which means that the magnetic $\mathrm{U}(1)$ symmetry is not broken. 
Thus, this phase corresponds to the unbroken phase of the dual complex scalar theory~(\ref{XY model}). 
%
%
The Abelian-Higgs model allows a (local) vortex solution, which carries an unit   charge under the unbroken magnetic $\mathrm{U}(1)$ symmetry. 
In the dual complex scalar theory (\ref{XY model}), this massive vortex state corresponds to the massive $\phi$ excitation in the unbroken phase. 
This is the reason why this duality is called Particle-Vortex duality.  

Although a rigorous proof of the duality has not been achieved, it is believed to be true by numerous studies in addition to numerical simulations~\cite{Nguyen_1999,Kajantie:2004vy}. 

\

As one can see from the above argument, the essence of the duality simply follows from the fundamental properties of the systems such as symmetry, spacetime dimension, and degrees of freedom, i.e. universality. 
Since we have formulated brane-field theory as a generalization of ordinary quantum field theory, it is tempting to see whether a similar IR duality exists in brane-field theories. 
%
%
Of course, a rigorous proof of such a duality would be much harder than the Particle-Vortex duality because we have to deal with the path-integral of a  functional field $\psi[C_p^{}]$ (not of a field $\phi(X)$), and this is beyond the scope of this paper.  

In the following, we conjecture a duality between the superconducting brane-field theory and its dual brane-field theory from the point of view of symmetry, order-parameters, and low-energy excitation modes.   
Since a dual object of brane is no longer particle or vortex in general, we call a new duality {\it Brane-Dual-Brane Duality}.

\subsection{Brane-Dual-Brane Duality Conjecture}
Our conjecture is as follows:  
The superconducting $p$-brane-field theory
\aln{\label{theory A}
Z_{\rm A}^{}[E_{d-p-2}^{}]&=\int {\cal D}\psi {\cal D} A~e^{iS_{\rm A}^{}}~,
\\
S_A^{}={\cal N}\int [dC_{p}^{}]\bigg\{-
\int_{\Sigma_d^{}}\frac{\delta(C_{p}^{})}{\mathrm{Vol}[C_{p}^{}]}&(D_{p+1}^{G}\psi[C_p^{}])^\dagger \wedge \star D_{p+1}^{G}\psi^{}[C_p^{}]-V(\psi^\dagger \psi)
\bigg\}-\frac{1}{2g^2}\int F_{p+2}^{}\wedge \star F_{p+2}^{}
\nn
&+\frac{1}{2\pi}\int_{\Sigma_{d}^{}}F_{p+2}^{}\wedge E_{d-p-2}^{}~.
\label{p brane AH}
}
is IR dual to the $(d-p-3)$-brane-field theory 
\aln{
\label{theory B}
Z_{\rm B}^{}[E_{d-p-2}^{}]&=\int {\cal D}\Psi e^{iS_{\rm B}^{}}~,
\\
S_B^{}={\cal N}\int [dC_{p}^{}]\bigg\{-
\int_{\Sigma_d^{}}\frac{\delta(C_{d-p-3}^{})}{\mathrm{Vol}[C_{d-p-3}^{}]}&(\tilde{D}_{d-p-2}^{}\Psi [C_{d-p-3}^{}])^\dagger\wedge \star \tilde{D}_{d-p-2}^{}\Psi^{}[C_{d-p-3}^{}]-\tilde{V}(\Psi^\dagger \Psi)\bigg\}~,
\\
\tilde{D}^{}_{d-p-2}\Psi[C_{d-p-3}^{}]=D_{d-p-2}^{}&\Psi[C_{d-p-3}^{}]-iG_{d-p-2}^{}\Psi[C_{d-p-3}^{}]~. 
\label{dual theory}
}
Namely, we conjecture 
\aln{Z_{\rm A}^{}[E_{d-p-2}^{}]\simeq Z_{\rm B}^{}[E_{d-p-2}^{}]\quad \text{near an IR critical point,}
}
where $E_{d-p-2}^{}$ is a non-dynamical background field.  
In the following, we simply call Eq.~(\ref{theory A}) theory A and Eq.~(\ref{theory B}) theory B. 
%
%

A couple of comments are necessary here. 
First, since we have not discussed any quantum aspects (or renormalization group) of these brane-field theories, the meaning of ``IR critical point" is vague in the above statement. 
However, as we have discussed in the Particle-Vortex duality, the essence of duality would be simply determined by the fundamental properties of the  systems such as symmetry, spacetime dimension, and degrees of freedom, without referring any details of the models. 
We therefore assume that there exists a (nontrivial) IR critical point in these brane-field theories and try to convince the readers by presenting several  correspondences of symmetry, order-parameter, and low-energy excitation modes. 
  %
  
Second, the actions (\ref{p brane AH})(\ref{dual theory}) are actually not the most general ones allowed by higher-form (gauge) symmetry and derivative expansion. 
For example, we can also consider more general interactions such as 
\aln{\label{topology changing interaction}
\int [dC_p^1]\int [dC_p^2]\int [dC_p^3]\delta(C_p^1-C_p^2-C_p^3)\psi^\dagger[C_p^1]\psi[C_p^2]\psi[C_p^3]+{\rm h.c.}~,
}
in both theories, which represents the splitting or merging of branes\footnote{This interaction still preserves the $\mathrm{U}(1)$ $p$-form (gauge) symmetry due to the delta function but violates the global $\mathrm{U}(1)$ ($0$-form) symmetry $\psi[C_p^{}]\rightarrow e^{i\theta}\psi[C_p^{}]~,~\theta \in \mathbb{R}$, which corresponds to the non-conservation of the total number of branes. 
}.
Such interactions seem to alter the dynamics of the brane field significantly and IR behaviors too.   
In the following, we simply neglect these interactions and focus on the above  theories. 

Since we have already discussed theory A in Section~\ref{sec:superconductor}, let us discuss the duality from the viewpoint of phases of theory B.

\

\noindent {\bf UNBROKEN PHASE OF THEORY B}\\
In the unbroken phase of theory B, the expectation value of the brane-field indicates the area law~\cite{Hidaka:2023gwh}
\aln{\langle \Psi[C_{d-p-3}^{}]\rangle\sim \exp\left(-c{\rm Vol}[M_{d-p-2}^{}]\right)~
}
in the large volume limit of $C_{d-p-3}^{}$.  
Here, $c$ is a constant and $M_{d-p-2}^{}$ is the minimal surface with the boundary $\partial M_{d-p-2}=C_{d-p-3}^{}$. 
Since this is a gapped phase, the lowest excited state would be the state created by the massive brane-field $\Psi^\dagger[C_{d-p-3}]$, which is a $(d-p-3)$-dimensional spatial object on a time slice.  
Correspondingly, its world-volume is a $(d-p-2)$-dimensional timelike subspace $\tilde{D}_{d-p-2}^{W}$.  

In theory A, this phase corresponds to the superconducting phase. 
The unbroken $\mathrm{U}(1)$ symmetry in theory B corresponds to the magnetic $(d-p-3)$-form symmetry $\mathrm{U}_{\rm M}^{}(1)$, and the brane-field $\Psi[C_{d-p-3}^{}]$ corresponds to the 't Hooft operator 
\aln{T[C_{d-p-3}^{}]=\exp\left(i\int_{C_{d-p-3}^{}}\tilde{A}_{d-p-3}^{}\right)~,
\label{'t Hooft operator}
}
where $\tilde{A}_{d-p-3}^{}$ is the dual field of $A_{p+1}^{}$.
This 't Hooft operator also indicates the area law because of the confinement of the magnetic flux as we have discussed in the previous section.   
Besides, the topological defect $D_{d-p-2}^W$ obtained in Section~\ref{sec:superconductor}  corresponds to the lowest excited state $\tilde{D}^W_{d-p-2}$ in theory B. 

\

\noindent {\bf BROKEN PHASE OF THEORY B}\\
In the broken phase of theory B, the expectation value of the brane-field shows  the perimeter law~\cite{Hidaka:2023gwh}
\aln{\langle \Psi[C_{d-p-3}^{}]\rangle\simeq v
\label{perimeter law}
}
in the large volume limit of $C_{d-p-3}^{}$, where $v$ is a VEV determined by  the potential $\tilde{V}(\Psi^\dagger \Psi)$.  
The Nambu-Goldstone mode is given by the phase modulation 
\aln{\Psi[C_{d-p-3}^{}]=v\exp\left(i\int_{C_{d-p-3}}A_{d-p-3}^{}\right)~,
\label{NG d-p-3}
}
and the low-energy effective theory is the $(d-p-3)$-form Maxwell theory~\cite{Hidaka:2023gwh}
\aln{
S_{\rm eff}^{}=-\frac{v^2}{2}\int_{\Sigma_d^{}}F_{d-p-2}^{}\wedge \star F_{d-p-2}~,\quad F_{d-p-2}^{}=dA_{d-p-3}^{}~. 
}
Furthermore, we can construct a topologically non-trivial static solution (soliton) which extends in a $p$-dimensional subspace and whose topological charge is given by
\aln{
Q_{p+1}^{}=\frac{1}{2\pi v^2}\int_{C_{d-p-2}}J_{d-p-2}^{}~,
\label{p+1 charge}
}  
where $C_{d-p-2}^{}$ is a closed subspace and 
\aln{
J_{d-p-2}^{}={\cal N}\int [dC_{d-p-3}^{}]\frac{\delta(C_{d-p-3}^{})}{{\rm Vol}[C_{d-p-3}^{}]}i(\Psi^\dagger  D_{d-p-2}^{}\Psi-\Psi(D_{d-p-2}^{}\Psi)^\dagger)~
\label{d-p-2 current}
}
is the $(d-p-2)$-form current. 
In fact, the static solution should approach a configuration like Eq.~(\ref{NG d-p-3}) in the spatial infinity and Eq.~(\ref{p+1 charge}) becomes 
\aln{
Q_{p+1}^{}=\frac{1}{2\pi}\int_{C_{d-p-2}}dA_{d-p-3}^{}=\frac{1}{2\pi}\int_{C_{d-p-2}}F_{d-p-2}^{}~,
}
which is an integer because of the Dirac quantization condition.  
For example, when $d=3$ and $p=0$, the scalar field behaves as $\Psi(X)\sim ve^{in\theta}~(n\in \mathbb{Z})$ for spatial infinity, and this leads to  
\aln{Q_1^{}=\frac{n}{2\pi}\int_{S^1}d\theta =n\in \mathbb{Z}~,
}   
which is nothing but the winding number of the global vortex. 

\begin{table}[t]
\hspace{-1cm}
\begin{tabular}{|c||c|c|c|c|}\hline
      & Order Parameter  &  Massless modes  & Massive modes         
            \\\hline\hline
Theory A ( Coulomb Phase ) & Perimeter law of $T[C_{d-p-3}^{}]$  &   $A_{p+1}^{}$     &   $\psi[C_{p}^{}]$ excitations  
 \\\hline
Theory B (Broken Phase) &  Perimeter law of $\Psi[C_{d-p-3}^{}]$  &  $A_{d-p-3}^{}$  &  $p$-dim solitons      
 \\\hline
 \end{tabular}
\caption{Duality between theory A in the Coulomb phase and theory B in the broken phase.  
This corresponds to a gapless phase. 
}
\label{tab:coulomb}
\end{table}
\begin{table}[!t]
\hspace{-1cm}
\begin{tabular}{|c||c|c|c|c|}\hline
      &  Order Parameter   & Massive modes          
            \\\hline\hline
Theory A (Superconducting Phase) &  Area law of $T[C_{d-p-3}^{}]$     &   $(d-p-3)$-dim topological defects  
 \\\hline
Theory B (Unbroken phase)  & Area law of  $\Psi[C_{d-p-3}^{}]$  &  $\Psi[C_{d-p-3}^{}]$ excitations    
 \\\hline
 \end{tabular}
\caption{Duality between theory A in the superconducting phase and theory B in the unbroken phase. 
This corresponds to a gapped phase. 
}
\label{tab:superconducting}
\end{table}

Now let us see how these physics can be interpreted in theory A.  
In theory A, this phase corresponds to the Coulomb phase. 
The perimeter law (\ref{perimeter law}) of $\Psi[C_{d-p-3}^{}]$ corresponds to the perimeter law of the 't Hooft operator (\ref{'t Hooft operator}), which  results in the spontaneous breaking of $\mathrm{U}_{\rm M}(1)$.    
The Nambu-Goldstone mode (\ref{NG d-p-3}) corresponds to the $(p+1)$-form gauge field $A_{p+1}^{}$ because they are dual-forms each other. 
In particular, the number of physical degrees of freedom is~\footnote{
This can be derived as follows~\cite{Hidaka:2020ucc}. 
The original (off-shell) degrees of freedom of the field strength $F_{p+2}^{}=dA_{p+1}^{}$ is $_{d}C_{p+2}^{}$. 
However, the Bianchi identity $dF_{p+2}^{}=0$ gives the $_{d-2}C_{p+2}^{}$ constrains and the EOM $d\star F_{p+2}^{}=0$ gives the  $_{d-2}C_{p}^{}$ constrains. 
Thus, the total number of physical degrees of freedom is
\aln{\frac{1}{2}\left(_{d}C_{p+2}^{}-{}_{d-2}C_{p+2}^{}-{}_{d-2}C_{p}^{}\right)={}_{d-2}^{}C_{p+1}^{}~,
}
where the factor $\frac{1}{2}$ comes from the fact that the physical modes consist of pairs of these degrees of freedom.
}
\aln{
{}_{d-2}^{}C_{p+1}^{}={}_{d-2}^{}C_{d-p-3}^{}~,
} 
Finally, a $p$-dimensional topological soliton in theory B corresponds to the massive excited state created by $\psi^\dagger[C_{p}^{}]$ in theory A.  

In Tables~\ref{tab:coulomb} and \ref{tab:superconducting}, we summarize the duality. 
Although there is no rigorous proof yet, the correspondences of symmetry, order-parameters, and degrees of freedom seem convincing enough to believe that this duality would be true.  
We leave more dedicated studies in future investigations.

\section{Summary}\label{Summary and Discussion}
We have proposed a field theory of $p$-branes interacting with a $(p+1)$-form gauge field $A_{p+1}^{}$ and studied its fundamental phenomena such as the Meissner effect, topological defect, and low-energy effective theory.  
We have found that all these phenomena are naturally and consistently extended to higher-dimensional branes, supporting the fact that our brane-field theory is a correct genralization of ordinary quantum field theory for particles to branes.    

Based on our brane-field theories, we have also conjectured a duality between the superconducting brane-field theory and dual brane-field theory with a global $\mathrm{U}(1)$ higher-form symmetry as a generalization of the Particle-Vortex duality. 
%
The clear correspondence between symmetry, order-parameters, and low-energy excitation modes provide us with sufficient reasons to believe in this generalized duality from the point of view of universality.  
We want to come back to this subject in the near future.

\section*{Acknowledgements}
We thank Yoshimasa Hidaka for the helpful discussions and comments. 
This work is supported by KIAS Individual Grants, Grant No. 090901.   
%


%
%
%

\bibliographystyle{TitleAndArxiv}
\bibliography{Bibliography}

\end{document}